\documentclass[prb,twocolumn,floatfix,groupedaddress]{revtex4}
\usepackage{graphicx}
\usepackage{amsmath}

\newcommand{\be}{\begin{equation}}
\newcommand{\ee}{\end{equation}}
\newcommand{\ben}{\begin{eqnarray}}
\newcommand{\een}{\end{eqnarray}}

\newcommand{\nd}{\noindent}
\begin{document}

\title{The thermodynamics of urban population flows}

\author{A. Hernando$^1$, A. Plastino$^{2,\,3}$}

\affiliation{
$^1$ Laboratoire Collisions, Agr\'egats, R\'eactivit\'e, IRSAMC, Universit\'e Paul Sabatier 118 Route de Narbonne 31062 - Toulouse CEDEX 09, France\\
$^2$ National University La Plata, Physics Institute (IFLP-CCT-CONICET) C.C. 727, 1900 La Plata, Argentina\\
$^3$ Universitat de les Illes Balears and IFISC-CSIC, 07122 Palma
de Mallorca, Spain}

\begin{abstract}
\nd Orderliness, reflected via  mathematical laws, is encountered
in different frameworks involving social groups. Here we show that
a thermodynamics can be constructed that macroscopically describes
urban population flows. Microscopic dynamic equations and
simulations with random walkers underlie the macroscopic approach.
Our  results might be regarded, via suitable analogies, as a step
towards building an explicit social thermodynamics.
\end{abstract}

\maketitle

\section{Introduction}

\nd The application of mathematical models to social sciences has
a long and distinguished history~\cite{1st}. One may speak of
empirical data from
scientific collaboration networks~\cite{cites}, cites of physics
journals~\cite{nosotrosZ}, the Internet traffic~\cite{net1}, Linux
packages links~\cite{linux}, popularity of chess
openings~\cite{chess}, as well as  electoral
results~\cite{elec1,ccg}, urban
agglomerations~\cite{ciudad,ciudad2} and firm sizes all over the
world~\cite{firms}. A specially relevant issue is that of
universality classes defined by to the so-called Zipf's law (ZL)
in the cumulative distribution or rank-size distributions
~\cite{zipf,nosotrosZ,net1,linux,chess,ciudad,ciudad2,firms,chin,upf,citis}.
Maillart et al.~\cite{linux} have found that  links' distributions
follow ZL as a consequence of stochastic proportional growth. Such
kind of growth assumes that an element of the system becomes
enlarged proportionally to its size $k$, being governed by a
Wiener process. The class emerges from a condition of stationarity
(dynamic equilibrium)~\cite{citis}. ZL also applies for processes
involving either self-similarity~\cite{chess} or fractal
hierarchy~\cite{chin}, all of them mere examples amongst very
general stochastic ones~\cite{upf}. A second universality-class
was found by Costa Filho et al.~\cite{elec1}, who studied
vote-distributions in Brazil's electoral results. Therefrom emerge
multiplicative processes in complex networks~\cite{ccg}.
Such behavior ensues as well in i) city-population rank
distributions~\cite{nosotros}, ii) Spanish electoral
results~\cite{nosotros}, and iii) the degree distribution of
social networks~\cite{nosotros2}. As shown in Ref.
\cite{benford2}, this universality class encompasses Benford's
Law~\cite{benford1}. In the present vein, still another kind of
idiosyncratic distribution is often reported:  the log-normal
one~\cite{lnwiki}, that has been observed in biology (length and
sizes of living tissue~\cite{bio}), finance (in particular, the
Black and Scholes model \cite{black}), and firms-sizes. The latter
instance obeys Gibrat's rule of proportionate
growth~\cite{gibrat}, that also applies to cities' sizes.

\nd Together with geometric Brownian motion, there is a variety of
models arising in different fields that yield Zipf's law and other
power laws on a case-by-case
basis~\cite{ciudad,ciudad2,citis,mod1,exp,renorm}, as preferential
attachment~\cite{net1} and competitive cluster growth~\cite{ccg,nosotros2}
in complex networks, used to explain many of the scale-free
properties of social networks.
For instance, we may mention detailed realistic approaches in
urban modelling~\cite{ud,otros}, opinion dynamics~\cite{oppi}, and
electoral results~\cite{elec1,elec2}.
Of course, the renormalization group is
intimately related to scale invariance and associated techniques
have been fruitfully exploited in these matters (as a small sample
see \onlinecite{renorm,voteGalam}).

\nd It has been recently shown, in Ref.~\onlinecite{epjb2}, that a variational
principle based on MaxEnt can be successfully applied to  scale-invariant social systems.
Used in the present context, it allows for a classification of the  above cited behaviors on the basis of
inferences drawn from objective observables of the system. We had also shown \cite{emp} that including some dynamical information
in the variational scheme \cite{epjb1} one is able to reproduce the shape
of empirical city-population distributions, going beyond the customary universality
classes conventionally used in such regards. Indeed, a connection between  explicit
microscopic growth equations and the macroscopic
characterization exists, illustrated for logistic-growth
in Ref.   \onlinecite{logistic}. We will here describe the manner in which the methods of that paper
can be generalized to first-principles theoretical framwework  describing  population flows in terms of thermodynamic concepts.

\subsection{Motivation, statement of the problem and goal}

\nd We are looking here for more that models: what we aim for is
to discover physical principles that may underlie some social
phenomena. Our system is a specific geographical area whose population
is distributed amongst several population-nuclei (cities, villages, towns, etc.)
Each nucleus' population is time-dependent due to migration, birth, death,
etc. Our aim is to quantitatively describe the population-nuclei's
variation. Microscopic variables are plentiful, but our main goal
is to be able to identify macroscopic variables that can give a
reasonable account of urban population-variations.

\vspace{0.2cm} \nd We will proceed in seven steps, as indicated in
the scheme below:\\
\fbox{\parbox{0.97\linewidth}{
\begin{enumerate}
\item Introduce the basic observables and the empirical data sets.
\item Identify the stochastic nature of the city-population growth rates.
\item Postulate dynamic microscopic equations and empirically validate them.
\item Perform numerical simulations with random walkers following these dynamical
      equations and parametrize the macroscopic evolution.
\item Show that equilibrium configurations of such evolutions can be predicted by
      MaxEnt using few macroscopic parameters.
\item Derive thermodynamic-like relations between these macro-parameters.
\item Show the applicability of our thermodynamic description by modeling empirical
      urban flows as an scale invariant ideal gas.
\end{enumerate}
}}

\nd The paper is organized as follows. Step 1 is addressed in the next
Section II. Section III deals with step 2, Section
IV with step 3, Section V with step 4, and Section VI with step 5 and 6.
Finally, the application is dealt with in Section VII, and some
conclusions are drawn in Section VIII.

\section{Preliminary matters}

\nd The basic ingredients we need in our approach, following Refs.~\onlinecite{emp,epjb1}, are
\begin{itemize}
 \item[i)]   $n$, the total number of ``population-nuclei";
 \item[ii)]  $x_i(t)$, the population of the $i$-th nucleus at time $t$
             (and $\mathbf{x}(t)=\{x_i(t)\}_{i=1}^n$ a vector with all the populations);
 \item[iii)] $x_0$ and $x_M$, the minimum and maximum allowed nucleus' population (in
             general $x_0=1$ and $x_M=\infty$);
 \item[iv)]  $N_T$, the total area's population ($N_T=\sum_{i=1}^nx_i(t)$);
 \item[v)]   $\dot{x}_i(t)$, the time-derivative of  $x_i(t)$ (thus the pairs
             $\{(x_i,\dot{x}_i)\}_{i=1}^n$ compose the ``urban phase space"); and
 \item[v)]   some a priori knowledge of the dynamics at hand, written as
             \be\label{dyn0}
             \dot{x}_i(t) = k_i(t) g_i[\mathbf{x}(t)]
             \ee
             where $g_i$ are population-functions to be determined and $k_i(t)$ growth rates
             independent of the $g_i$.
\end{itemize}

\nd  The raw data used in our analysis is obtained from the Spanish state institute
INE\cite{ine} and cover annually the period 1996-2010 (with the exception of
1997). It encompasses up to 8000 municipalities (the smallest
Spanish administrative unit) distributed within 50 provinces (the
building blocks of the autonomous communities). We use provinces
and municipalities as the closest representatives of the ideal of a
closed system's fundamental elements. Also other regions of the world are
used as examples along the text. In this tableau, the total
population $N_T$ of a province is apportioned in $n$ nucleus.
The $i$-th nucleus account a population of $x_0\leq x_i(t)\leq x_M$ at time $t$,
which time-evolution obeys Eq.~(\ref{dyn0}).

\section{The stochastic nature of population growth rates}

\begin{figure}
\includegraphics[width=0.45\textwidth]{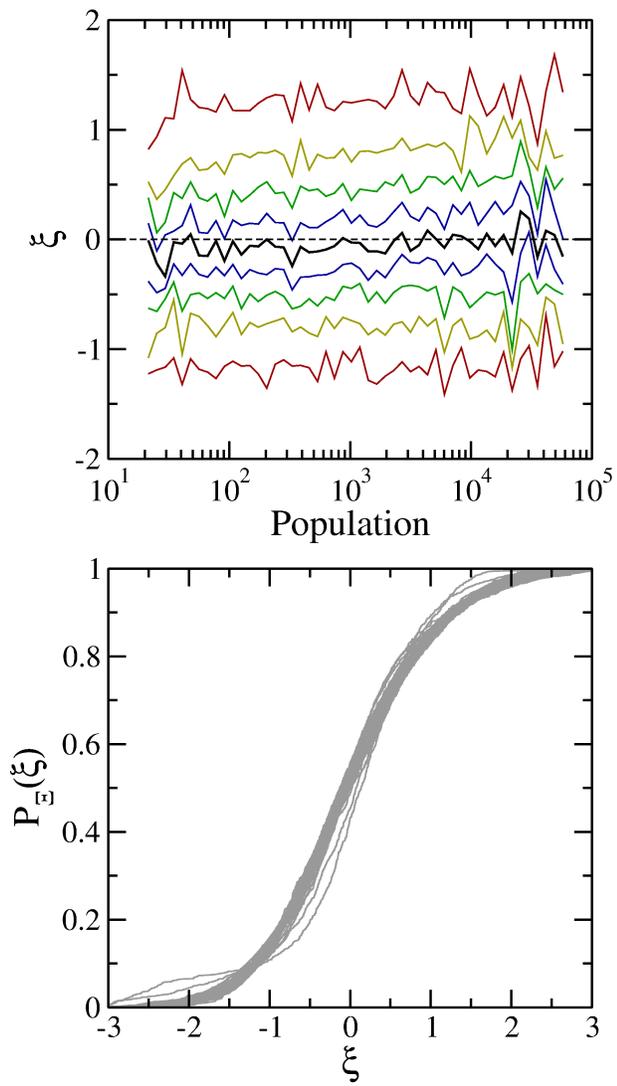}
\caption[]{Top panel: quantiles from 0.1 to 0.9 each 0.1
for $p_\Xi(\xi)$ as a function of the population $x$
(the median in shown in black).
Bottom panel: $P_\Xi(\xi)$'s cumulative
distribution for each of Spain's provinces}.\label{fig1}
\end{figure}

\nd We begin dealing with step 2 of our Scheme, saying something
meaningful concerning the form of the growth rates $k_i(t)$ in
Eq.~(\ref{dyn0}). The value of $k_i$ above depends upon millions of
individual decisions, so it is expected some stochastic behavior. We
should know both the average $m_i=\langle\dot{x}_i(t)\rangle_{\delta t}$
and the standard deviation $s_i=\langle(\dot{x}_i(t)-m_i)^2\rangle_{\delta t}^{1/2}$
(for each $i$) in a time-window ${\delta t}$ around $t$. Trying then to study the distribution of
$\xi_i(t) = (\dot{x}_i(t)-m_i)/s_i$ one immediately finds
\ben
m_i   &=& \langle k_i(t)g_i(t)\rangle_{\delta t}\nonumber\\
      &=& \langle k_i(t)\rangle_{\delta t}\times \langle g_i(t)\rangle_{\delta t},\\
s_i^2 &=& \left\langle (k_i(t)g_i(t)-\langle k_i(t)g_i(t)\rangle_{\delta t})^2 \right\rangle_{\delta t}\nonumber\\
      &=& \sigma_{k_i}^2 \langle g_i(t)\rangle_{\delta t}^2-\langle k_i(t)\rangle_{\delta t}^2~\sigma^2_{g_i}
\een
with $\sigma_{k_i}$, $\sigma_{g_i}$ being the standard deviations
of $k_i(t)$ and $g_i(t)\equiv g_i[\mathbf{x}(t)]$, respectively. Assuming
now that the function $g_i$'s variation in the time-window for which one
evaluates  the pair $m_i$ - $s_i$ is negligible (i.e.,
$\sigma_{g_i}^2\ll\sigma_{k_i}^2$),  to a good approximation one has
\be
\xi_i(t) = \frac{k_i(t)-\langle k_i(t)\rangle_{\delta t}}{\sigma_{k_i}},
\ee
entailing that $\xi_i(t)$ has null average and unit standard deviation.
If this assumption is correct the shape of the $p_\Xi-$distribution of
the variable $\xi_i(t)$ should not depend upon  $x_i(t)$. We have verified
the hypothesis, {\it as our first result here}, with reference to all
(8116) Spain's municipalities. Fig.~\ref{fig1} displays the $(x_i,\xi_i)-$pairs
for every township in the time-window $\delta t=15$ years. From them we evaluate appropriate points taken
at regular intervals from the cumulative distribution function of
our random variable (quantiles) as a function of the population
$x$. No apparent $x-$dependence can be detected. The overall distribution
$p_\Xi(\xi)$ shape looks like a normal one
\be
p_\Xi(\xi) = \frac{e^{-\xi^2/2}}{\sqrt{2\pi}},
\ee
with cumulative distributions of the form
\be
P_\Xi(\xi) = \frac{1}{2}\left[1+\mathrm{erf}\left(\frac{\xi}{2}\right)\right],
\ee
shown in Fig.~\ref{fig1}. Save for some fluctuations, we have not found
any dependence on the shape of $p_\Xi(\xi)$ for the different  provinces (same
Fig.~\ref{fig1}). Accordingly, but with a grain of salt, one may speak of
``universality". Consequently, we will consider herefrom  that our variable $\xi$
can be regarded as belonging to a Wiener process, {\it our second
result here}.

\section{Introducing microscopic equations of motion}

\subsection{Proportional growth}

\begin{figure}
\includegraphics[width=0.45\textwidth]{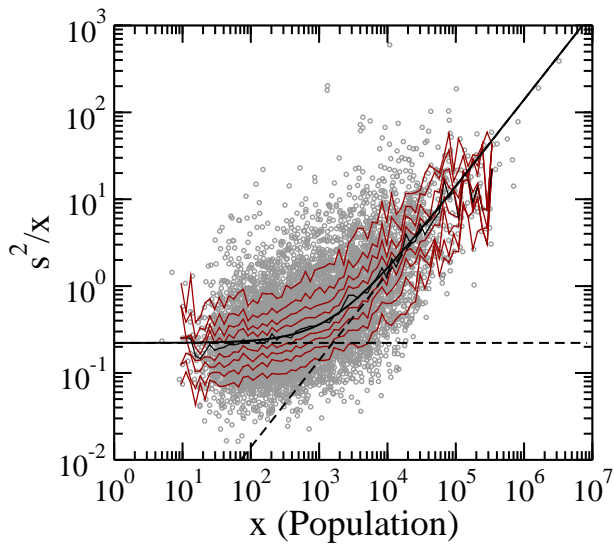}
\caption[]{Variance $s^2/x$ vs. $x$. Red: quantiles from 0.1 to 0.9 each 0.1.
Solid black: fit to the median value. Dashed black lines: Finite-size's fluctuations
are constant, while the multiplicative regime is given by  a straight line.}\label{fig2}
\end{figure}

\nd We are now at step 3. For the $g_i$'s shape we will assume that it
depends only on its own $x_i$'s population, i.e.,
$g_i[\mathbf{x}(t)]\simeq g_i[x_i(t)]$. In order to guess the explicit
analytical form we appeal to a cluster-growth model in networks,\cite{ccg,nosotros2} used
successfully before describing city-population distributions. We
firstly consider a network of nodes (that eventually represents the social
network) and a single node as seed of a cluster. Initially, the first
neighbors of the seed will belong to the cluster with a given probability
$P(t=0)$. At a subsequent time $t$, the first neighbors of the members of the
cluster become also members with probability $P(t)$. Proceeding in this vein,
it is reasonable to conjecture that the time-variation of the cluster-size $\dot{x}$ at time $t$ acquire the  form
\be
\dot{x}(t) = \sum_{j=1}^{x(t)}P(t) c_j(t),
\ee
where $c_j(t)$ is the first-neighbors-number of node $i$-th at time $t$.
We appeal now to the \emph{central limit theorem} to write
\be
\dot{x}(t) = P(t)\left(\overline{c}(t)x(t)+\sigma_c(t)\sqrt{x(t)}\xi(t)\right).
\ee
Here $\overline{c}(t)$ is the mean neighbor's number at time $t$, $\sigma_c(t)$ its standard
deviation, and $\xi(t)$ an independent normally-distributed number. This last summand, usually
neglected for very large sizes, is associated to finite-size effects. The first
term, size-proportional, generates proportional (or multiplicative) growth.  In
view of this result, we consider the form
\be
g_i(x_i) = [x_i]^\alpha
\ee
with $\alpha=1$ or $1/2$. Considering then both terms in the microscopic dynamics
we write
\be
\dot{x}_i(t) = k_{i1}(t)x_i(t) + k_{i1/2}(t)\sqrt{x_i(t)},
\ee
with the $k_{i1}(t)$ and $k_{i1/2}(t)$ two (a priori) independent
Wiener coefficients. This dependence is checked out by
comparison of the previously employed $s_i-$numbers with a
functional form of the type
\ben \label{recastol}
s_i^2(x_i) &=& \langle [\dot{x}_i]^2-\langle \dot{x}_i\rangle_{\delta t}^2 \rangle_{\delta t}\nonumber\\
           &=& \sigma_{i1}^2 x_i^2 + \sigma_{i1/2}^2 x_i,
\een
where $\sigma_{i1}$ and $\sigma_{i1/2}$ are the associated
deviations of $k_{i1/2}$ and  $k_{i1}$, respectively.
Rewriting (\ref{recastol}) in a more convenient way we have
\be
s_i^2(x_i)/x_i = \sigma_{i1}^2 x_i + \sigma_{i1/2}^2,
\ee
that, for sizes small enough reduces to
\be
s_i^2(x_i)/x_i \approx \sigma_{i1/2}^2,
\ee
while for very large sizes one has
\be
s_i^2(x_i)/x_i \approx \sigma_{i1}^2x_i.
\ee
\nd The transition between these two regimes should take place at
a value $x_T=\sigma_{1/2}^2/\sigma_{1}^2$. Fig.~\ref{fig2}
displays, as {\it our third result}, the $(x_i,s_i^2(x)/x_i)-$pairs
for all the Spanish municipalities, together with appropriate quantiles.
The median $\mathrm{med}(s_i(x_i)/x_i)$ nicely fits things with $\sigma_{i1}
= 0.0119$ and $\sigma_{i1/2}=0.47$. We appreciate the fact that
finite size fluctuations are larger than multiplicative ones, the
later dominating, of course, for large sizes. Our transition
occurs at population-values of the order of $1500$ inhabitants.
Surprisingly enough, the distribution of the variable
$s'_i=\log[s_i(x_i)/\sqrt{x_i}]-\log[\mathrm{med}(s_i(x_i)/\sqrt{x_i})]$
becomes independent of $x_i$, being of a Gaussian nature.

\nd At this point, we need still to address a further question.
The finite-size term average is $\langle k_{i1/2}(t)\rangle_{\delta t}=0$ (by definition),
but this is not so for the multiplicative one $\langle k_{i1}(t)\rangle_{\delta t}\neq0$,
that is \emph{a priori} regarded as constant and size-independent.
This is indeed empirically true on occasions, but not always. For instance,
such assumption cannot account for the migration from the countryside
to big cities, where the mean growth rate correlates with the city-population.

\subsection{Taking into account internal flow}

\nd It is a fact that small populations tend to diminish while
large towns tend to increase their population. We encounter this
scenario for most of the 50 provinces of Spain.  We intend to
tackle this issue below. \vskip 2mm

\begin{figure*}
\includegraphics[width=0.95\textwidth]{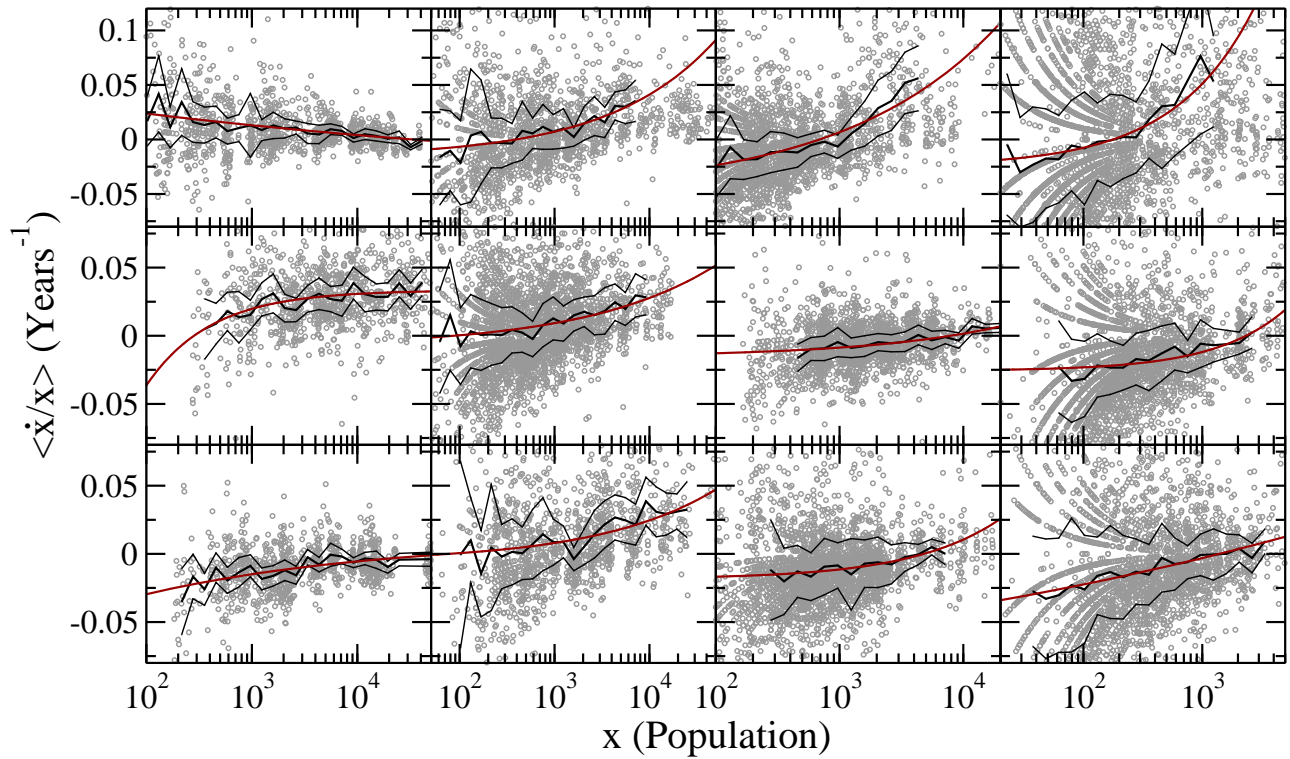}
\caption[]{Fit of $\langle\dot{x}/x\rangle$ to an expression of the
type represented by  Eq.~(16) for 12 Spanish provinces:
Asturias, Almer\'ia, C\'aceres, Cuenca, Baleares, Lleida,
Badajoz, \'Avila, Guip\'uzcoa, Castell\'on, Valladolid
and Guadalajara.
 }\label{fig3}
\end{figure*}

\nd We can show that the effect can be described by recourse to a
smooth dependence of the mean relative growth $\langle \dot{x}_{i}/x_i\rangle$
on $\log(\langle x\rangle)$ that generates what we will call internal flow. A second
order expansion in $\log(\langle x\rangle)$ reads
\be
\langle \dot{x}/x\rangle \simeq a + b\log(\langle x\rangle) + c\log(\langle x\rangle)^2
\ee
where the values of $a$, $b$ and $c$ come from the corresponding
Taylor coefficients. Assuming $b \gg c$ we can safely write it as
\be\label{dotx}
\langle \dot{x}/x\rangle \simeq \langle k_{1}\rangle + \langle k_{q}\rangle[\langle x\rangle]^{q-1},
\ee
where we have defined for convenience $\langle k_{1}\rangle=a-b^2/2c$,
$\langle k_{q}\rangle=b^2/2c$ and $q-1=2c/b$. To validate our assumptions,
we fitted the empirical provincial data to Eq. (\ref{dotx}) via $\langle k_{1}\rangle$,
$\langle k_{q}\rangle$ and $q$, when possible (in some cases a quasi-linear relation
is found, generating large uncertain in the optimal values). We have found
for the exponent $q$ a mean value of 1.2 and a standard deviation of 0.45,
with $|q-1|<1$ in all cases. This result confirms the assumption $b \gg c$
validating the second-order expansion of $\langle \dot{x}_{i}/x_i\rangle$.
Moreover, as seen in Fig.~\ref{fig3} ({our fourth result}), nice fits are
found in general with very few exceptions.

\nd With this new hypothesis our complete dynamic equation turns out to be
\be
\dot{x}_i(t) = k_{iq}(t)[x_i(t)]^q + k_{i1}(t)x_i(t) + k_{i1/2}(t)\sqrt{x_i(t)},
\ee
with $k_{iq}(t)$, $k_{i1}(t)$ and $k_{i1/2}(t)$ independent (a priori)
Wienner processes. Summing up, we have assumed
\begin{itemize}
  \item a finite size term that dominates things for low population levels ($<1500$),
  \item a multiplicative term that accounts for population's
  growth/diminution (births, death or o external migration, and
  \item a power-law (exponent $q\sim1$) accounting for internal migration.
\end{itemize}
Since for most of the population range only one term dominates,
 we will include only one term in the considerations what follow below.

\section{From microscopic to macroscopic descriptions}

\nd We arrive to stage 4, having discussed above a microscopic
population dynamics. We will tray now to ascertain whether a
{\it macroscopic} description is also feasible.
Our goal is to reduce the $2n$ microscopic degrees of freedom to a
few macroscopic ones. We will separately consider each of the three
terms of the dynamic equation. The ensuing results will be valid in
the domains in which each term dominates.

\nd Consider $n$ random walkers characterized by a dynamic coordinate
$x_i(t)$ obeying
\be\label{dyn}
\dot{x}_i(t) = k_i(t)[x_i(t)]^q,
\ee
with $\langle (k_i(t)-\overline{k})(k_j(t)-\overline{k})\rangle=\sigma_k\delta_{ij}\delta(t-t')$.
Parameter $q$ will take as special possible values $1/2$ or $1$, or
in general, $0\leq q$.

\subsection{Brownian motion and diffusion equation}

\nd We start with $q=0$ as control case. One has $\dot{x}_i(t) = k_i(t)$ so
that we deal with the well-known brownian random walkers. Consider this numerical
procedure: initially, the $n$ walkers are located at, say, $x=x_0$. By
$\rho(x,t)dx$ we will refer to the walker's normalized histogram, at time
$t$, that indicates the walker's relative number positioned in the interval
$dx$ around $x$. The associated initial density would read
$\rho(x,0)=\delta(x-x_0)$. A discrete version of the pertinent
dynamic equation is
\be
x_i(t+\Delta t) = x_i(t) + \Delta t k_i(t),
\ee
that forces the walkers to ``move" during the period $\Delta t$ in a
amount given by $\Delta t k_i(t)$, with $k_i(t)$ a random number generated
from a Gaussian distribution determined by an standard deviation $\sigma_k$
and mean $\overline{k}$, as defined above. We have
\be
x_i(t=M\Delta t) = x_0 + \Delta t\sum_{m=1}^{M} k_i[(M-1)\Delta t],
\ee
so that after $M$ iterations the walkers-distributions coincides with that
of a random number generated by summing up $M$ Gaussian numbers characterized
by $\Delta t\sigma_k$ and $\Delta t\overline{k}$. Remind that a distribution that
follows a random number composed of two other numbers of that
character is the convolution of the distributions associated to
these later numbers. Thus, $x(t)$ is described by the $M-$th
convolution of the $k$'s Gaussian distribution. By recourse to a
Fourier transform $\mathcal{F}$ for convolutions we have
\ben
\mathcal{F}[\rho(x,t)] &=&
\left(\mathcal{F}\left[\frac{e^{-(k-\Delta t\overline{k})^2/2(\Delta t\sigma_k)^2}}{\sqrt{2\pi}\Delta t\sigma_k}\right]\right)^M\nonumber\\
 &=& \left(e^{-\Delta t^2\sigma_k^2\omega^2/2 + i\Delta t\overline{k}\omega}\right)^M\nonumber\\
 &=& e^{-M\Delta t^2\sigma_k^2\omega^2/2 + iM\Delta t\overline{k}\omega},
\een
and, appealing to the inverse transformation,
\ben
\rho(x,t) &=& \frac{e^{-(k-M\Delta t\overline{k})^2/(2M(\Delta t\sigma_k)^2)}}{\sqrt{2\pi M}\Delta t\sigma_k}\nonumber\\
 &=& \frac{e^{-(k-t\overline{k})^2/(4Dt)}}{\sqrt{4\pi Dt}},
\een
where we have introduced for convenience $2D=\Delta t\sigma_k^2$.
An arbitrary density $\rho(x,t)$ will evolve in
$\Delta t$, via the convolution of that density with a Gaussian of
deviation $\Delta t\sigma_k=\sqrt{2\Delta tD}$ and mean $\Delta t\overline{k}$, as
\ben
\mathcal{F}[\rho(x,t+\Delta t)] &=& \mathcal{F}[\rho(x,t)]\times e^{-\Delta tD\omega^2 + i\Delta t\overline{k}\omega}\\
 &\simeq& \mathcal{F}[\rho(x,t)]\left( 1-\Delta tD\omega^2 + i\Delta  t\overline{k}\omega\right),\nonumber
\een
where we take  $\Delta t$ arbitrarily small. A simple
manipulation involving division by $\Delta t$ leads now to
\be
\frac{\mathcal{F}[\rho(x,t+\Delta t)] - \mathcal{F}[\rho(x,t)]}{\Delta t} =
\left(-D\omega^2 + i\overline{k}\omega\right)\mathcal{F}[\rho(x,t)].
\ee
By recourse to the inverse transformation and taking the limit $\Delta t\rightarrow0$ we get
\be\label{dif}
\partial_t \rho(x,t) = D\partial^2_x\rho(x,t) - \overline{k}\partial_x\rho(x,t),
\ee
which is a diffusion equation. {\sf Accordingly, we reach an
important result here (our {\it fifth one})}:\\
\nd \fbox{\parbox{0.97\linewidth}{
Our original $2n$ degrees of freedom-problem can now be tackled via just a few macroscopic parameters.}}

\subsection{$q$-metric Brownian motion}

\nd In the general instance $q\neq0$ we introduce a variable
$u_i=\log_q(x_i)$, where $\log_q$ is Tsallis' $q$-logarithm
\cite{tbook}. The Jacobian for the transform is $du/dx=1/x^q$ so
that $\dot{u}=\dot{x}/x^q$ and the associated dynamical equation
becomes
\be
\dot{u}_i(t) = k_i(t).
\ee
In the set $\{(u_i,\dot{u}_i)\}_{i=1}^n$, the variables $u_i$ and
$\dot{u}_i$ are independent of each other. We regard them, of course, as our dynamical
variables. Note that one recovers Brownian motion for $u$. Indeed,
\be
u_i(t=M\Delta t) = u_i(0) + \Delta t\sum_{m=1}^{M} k_i[(M-1)\Delta t],
\ee
and then the demonstration of the preceding subsection becomes valid,
now for $u$ and $\rho(u,t)du$. Our new diffusion equation reads
\be\label{difu}
\partial_t \rho(u,t) = D\partial^2_u\rho(u,t) - \overline{k}\partial_u\rho(u,t),
\ee
and, starting from a density $\rho(u,0)=\delta(u-u_0)$ we end up with
\be
\rho(u,t)du = \frac{du}{4\pi Dt}\exp\left[-\frac{(u-u_0-\overline{k}t)^2}{4Dt}\right].
\ee
The $x-$density is governed accordingly by a $q$log-normal distribution
\ben
\rho_X(x,t)dx &=& \rho[u(x),t]\frac{dx}{du}du\\
              &=& \frac{dx}{\sqrt{4\pi Dt}x^q}\exp\left[-\frac{(\log_q(x)-u_0-\overline{k}t)^2}{4Dt}\right].\nonumber
\een
In particular, for $q=1/2$ one has
\be \label{ln05}
\rho_X(x,t)dx = \frac{dx}{\sqrt{4\pi Dtx}}\exp\left[-\frac{(2(\sqrt{x}-1)-u_0-\overline{k}t)^2}{4Dt}\right],
\ee
and, for $q=1$ the well known log-normal
\be\label{ln}
\rho_X(x,t)dx = \frac{dx}{\sqrt{4\pi Dt}x}\exp\left[-\frac{(\log(x)-u_0-\overline{k}t)^2}{4Dt}\right].
\ee

\nd \fbox{\parbox{0.97\linewidth}{We have again reduced the microscopic
number of degrees of freedom to just a few macroscopic
parameters.}}


\subsection{Examples of diffusion}

\begin{figure}[ht]
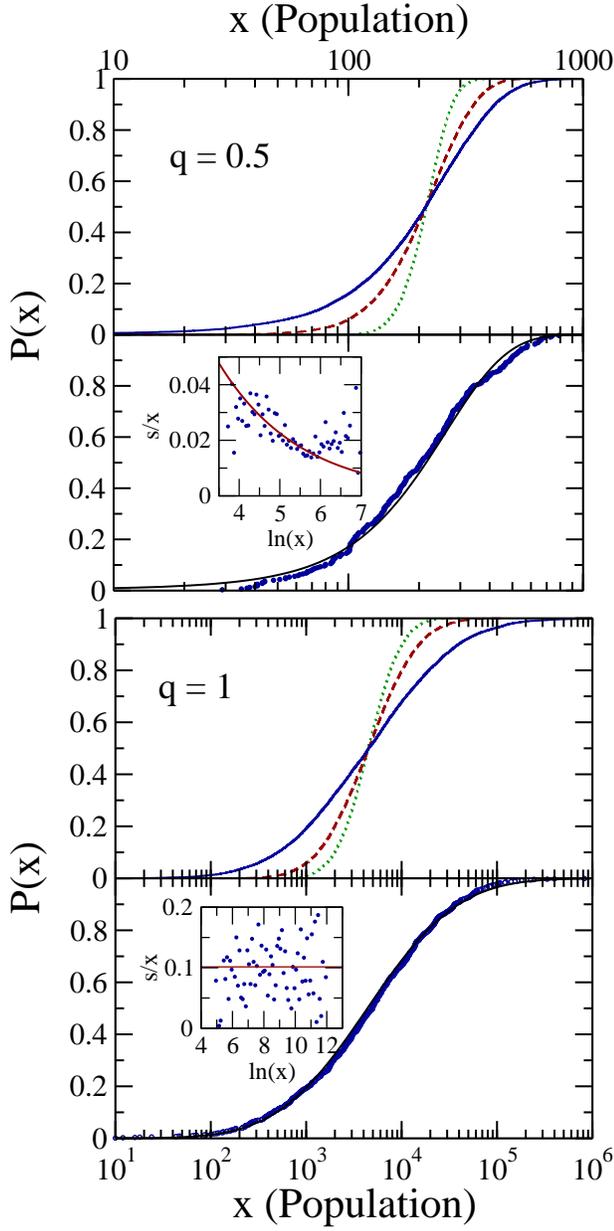

\includegraphics[width=0.45\textwidth,clip=true]{fig4b.eps}
\includegraphics[width=0.45\textwidth,clip=true]{fig4a.eps}
\caption[]{Top panel: $q=21/2$-metric diffusion at times $t=20$ (green dotted line), 70 (red dashed) and 110 (blue solid)
using the text-parameters compared with
the distribution of Salamanca towns' population in 2010, fitted to a $1/2$-log-normal 
distribution (solid line). Inset: variance of
the relative growth vs. log-population (dots), confirming the $\sqrt{x}$
dependence for $q=1/2$ dynamics (red line). Bottom panel:
geometric diffusion $(q=1)$ at times $t=4$ (green dotted), 9 (red dashed) and 29 (blue solid) compared with the population distribution
of Florida State (US) in 2010, fitted to a log-normal distribution (solid line). Inset: same as top panel's inset, confirming
 size independence and thus proportional dynamics.}\label{fig4}
\end{figure}

\nd Numerical experiments confirm our findings above. We start with our
dynamical equation in discrete form
\be\label{dyn1}
x_i(t+\Delta t) = x_i(t) + \Delta t k_i(t)[x_i(t)]^q
\ee
using $k_i(t) = \sqrt{2D/\Delta t}\xi_i(t)+\overline{k},$ where the
random numbers $\xi$ follow a normal distribution such that
$\langle\xi_i(t)\xi_j(t)\rangle=\delta_{ij}\delta(t-t')$. We have taken
$q=1/2$ and $1$ for our examples, and find that the associated
distributions exactly follow the diffusion equation's predictions.
We have used in the former case $u_0=\log_{1/2}(220)$, $\overline{k}=0$, and
$\sigma_k^2=10$, in intervals of $\Delta t=0.01$. In the later
instance we had $u_0=\log(4400)$ instead. Indeed, the walkers' histograms'
evolution follow Eq. (\ref{ln05}) and Eq. (\ref{ln}),
respectively, with $D=\Delta t\sigma_k^2/2$ as defined above (see
Fig.~\ref{fig4} for the cumulative distributions).

\nd As empirical examples we discovered that
for small populations $<1500$ inhabitants the finite-size noise dominates.
Provinces for which most towns are scarcely populated will obey the
dynamical equation with $q=1/2$. Such is the case for the province of,
i.e., Salamanca, as shown in top panel of Fig.~\ref{fig4}. The ensuing
dynamics confirms this assertion. The relative growth of most
of the
towns follows a dynamics with a variance  $s^2\propto \sqrt{x}$
(red line of the inset). The ensuing distribution fits the final
state predicted by the diffusion equation for that dynamics, Eq.
(\ref{ln05}),
with $u_0+\overline{k}t=\log_{1/2}(216.3)$ and $2Dt=95.6$ for year 2010 (see  Fig. 4).
Remark  that the 1/2-log-normal can be easily confused with the usual
log-normal, although the former exhibits asymmetries in log-scale. As a
$q=1-$example we mention Florida State in the US \cite{usa}
(see also bottom Fig.~\ref{fig4}). Using data from 1990, 2000, and 2010, we have
verified that the microscopic dynamics confirms the
proportional growth assumption (with a variance of the relative growth independent
of the size, as illustrated  in the inset). The city-populations distribution
follows a log-normal distribution, that of  Eq. (\ref{ln}), which can be the one
pertaining to geometrical random-walkers' diffusion,
with $u_0+\overline{k}t=\log(4380)$ and $2Dt=2.96$.

\subsection{Constrained diffusion}

\nd   $q$-log-normal distributions do not set any limits to
 population-sizes. However, it is reasonable to
assume that physical space does pose limits to a city's
population-growth. Unlimited growth is unrealistic since in the
case of internal migrations the total population $N_T$ should
remain constant and a free-diffusion model is, again,
unrealistic. Constrained diffusion must be contemplated instead.\\

\begin{figure*}
\includegraphics[width=0.95\textwidth]{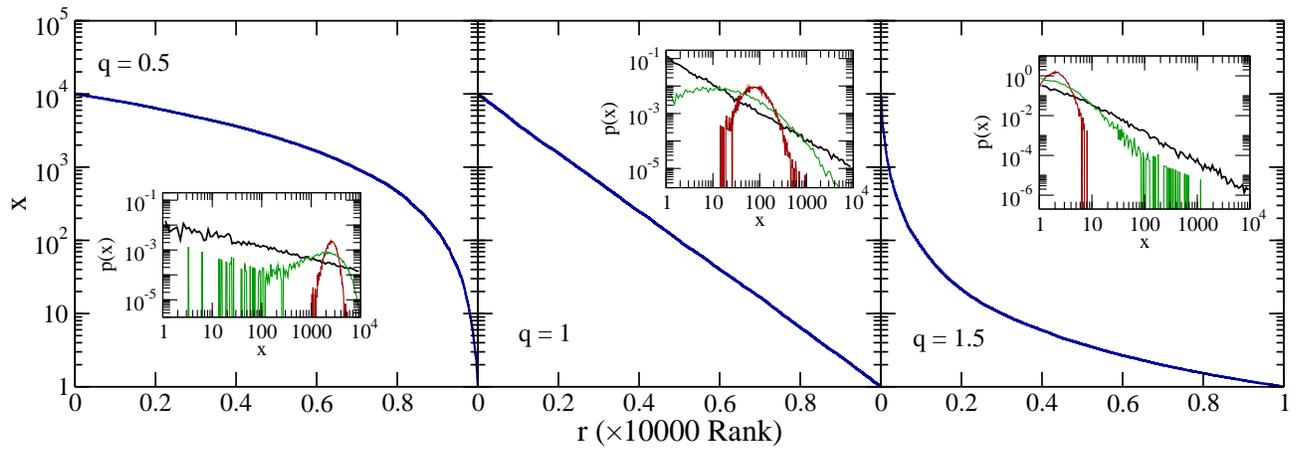}
\caption[]{$q$-metric diffusion with maximum size constraint for (from left to right panels)
$q=1/2$, $1$ and $1.5$. Rank-distributions and evolution (insets).}\label{fig5}
\end{figure*}

\nd We pass now to consider numerical experiments with random walkers
that fix lower and upper bounds for population. These are denoted
by $x_0$ and $x_M$, respectively. Now, walkers ``moves" leading to values
outside the range $x_0<x<x_M$ are to be rejected in our simulations.
Fig.~\ref{fig5} shows that a $q$-metric walkers' evolution begins by
faithfully following the diffusion equation Eq. (\ref{difu}) till they bump off these extreme values.
Now their density deviates from that of ``free" evolution. After
some time has elapsed, an equilibrium $x-$distribution is reached
that follows a power-law with exponent $q$, independently of the
initial state. \emph{The origin of this systematic result can not be
unraveled by the simulations, so a higher-level of theory is needed.}\\

\begin{figure}
\includegraphics[width=0.45\textwidth]{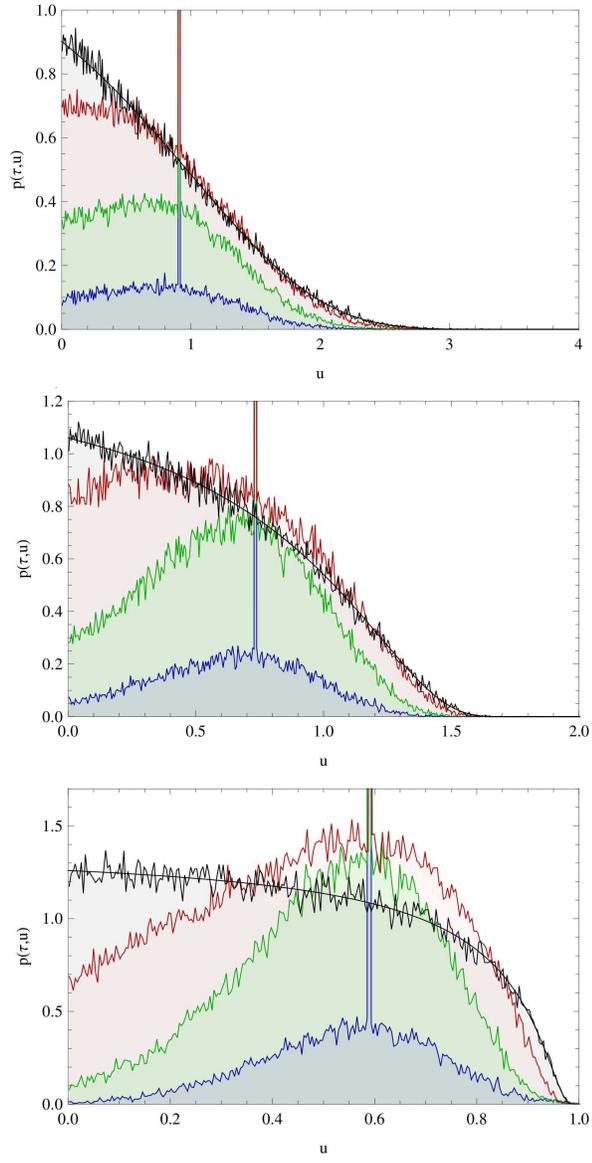}
\caption[]{$N$-constrained $q$-metric diffusion for $q=1$,
$1.5$, and $2$.}\label{fig6}
\end{figure}

\nd Now we use a total population constraint. This is equivalent to
make the walkers move under the rule of a \emph{$q$-generalized
multi-component logistic equation}
\be\label{dyn2}
\dot{x}_i(t) = [x_i(t)]^q\left[k_i(t)-\frac{\sum_{i=1}^nk_i(t)[x_i(t)]^q}{\sum_{i=1}^n[x_i(t)]^q}\right].
\ee
Indeed, it is easy to check that $\partial_t N_T=\sum_{i=0}^n\dot{x}_i(t)=0$,
thus preserving the value of $N_T$ in time. Also the original $q$-symmetry of
the dynamics is preserved. This equation is the $q$-generalization of the
scale-invariant multi-component logistic equation presented in \onlinecite{logistic}.
Results are displayed in Fig.~\ref{fig6} for $q=1$, 1.5 and 2,
using $n=100000$ walkers and a total population of $N=250000$
inhabitants (with $x_0=1$).
Remarkably enough, equilibrium is always
reached, to a density that does not depend upon the initial state or the $k$-parameters. The shape
of the distributions resembles  $x$ power-laws with exponential cut-off.
Again, the simulation can not unravel the origin of this form. \emph{Finding the
properties and the exact analytical form of those macroscopic
equilibrium distributions is our goal in the sext Section.}

\section{The macroscopic conundrum}

\nd  We tread now step 5. Our simulations with random walkers
suggest that it is indeed possible to pass from a description
that uses $2n$ microscopic variables to a description involving
just a few macroscopic parameters. The big question is: do they
behave in thermodynamic fashion, satisfying the pertinent partial
derivatives-relationships? We wish to tackle this issue now
looking for a way to reduce the number of microscopic degrees of
freedom to a few manageable macroscopic ones while keeping a
coherent, reasonable description of our system, mimicking the kind
of scenario that links statistical mechanics to thermodynamics.
This requires appropriate constraints, a topic to be addressed
below by enumerating the appropriate ``social" constraints we
need.

\subsection{Macroscopic constraints}

$\bullet$  {\it Total number of cities $n$.}
Since there is some confusion
in the available data about what the administrative meaning of city is,
we wish to ascertain that this issue is of no importance. Consider
$x_i=\sum_j^{n_i}x_{ij},$ where $n_i$ is the number of sub-administrative
units included in the administrative unit $i$, with $x_{ij}$ their
sub-administrative populations. Considering
proportional growth, we write for the time-evolution
\ben
\dot{x}_i(t)
&=&\sum_j^{n_i}\dot{x}_{ij}(t)\nonumber\\
&=&\sum_j^{n_i}k_{ij}(t)x_{ij}(t)\nonumber\\
&=&\frac{\sum_j^{n_i}k_{ij}(t)x_{ij}(t)}{\sum_j^{n_i}x_{ij}(t)}\sum_j^{n_i}x_{ij}(t)\nonumber\\
&=&k'_i(t)x_i(t),
\een
where we have defined $k'_i(t)$ as a new variable defined as an average
weighted by the populations $x_{ij}$. If the growth rates $k_{ij}$ are random
variables with approximately the same mean and variance, it is easy to check
that $k'_i(t)$ is in turn a random variable of the same mean and variance.
The dynamical behavior of the ensemble of
administrative units $\mathbf{x}$ is thus equivalent of that of the
sub-units, and the procedure described in this work is still applicable.\\

\begin{figure*}
\includegraphics[width=0.9\textwidth]{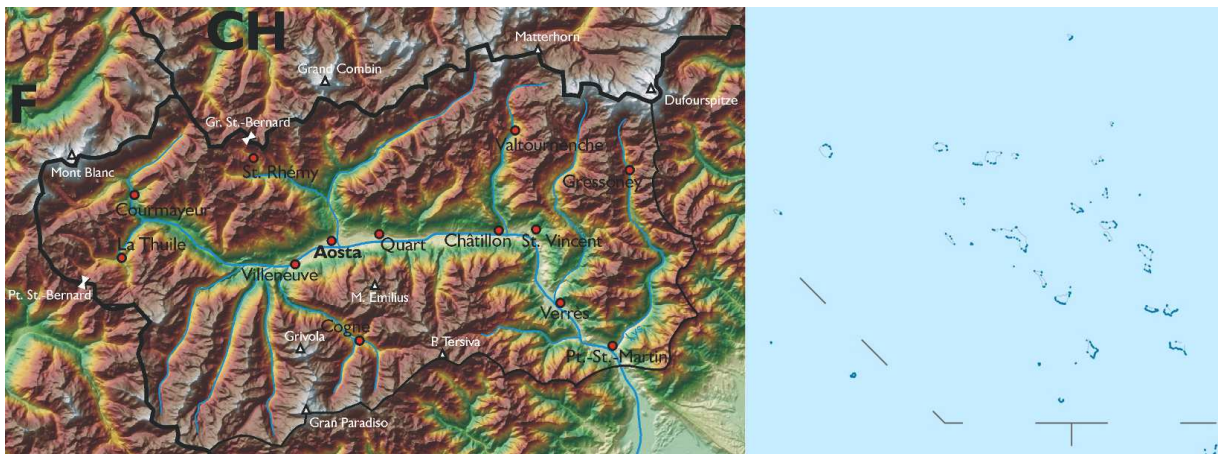}
\caption[]{An example of population restriction arising out of
geographical reasons. Left Panel:
d'Aosta valley (Italy)\cite{aostamap}. Right Panel: Marshall islands\cite{marmap}.}\label{fig8}
\end{figure*}

$\bullet$  {\it Maximum/minimum population $x_M/x_0$.}
It is well-known that a typical minimum population size equals the
Dunbar number\cite{dum} ($\sim 150$), heuristically associated to
the maximum (allowable by our neo-cortex) number of stable
  human relationships. Thus, it is reasonable to think of a
minimum size $x_0\sim150$. In many cases a maximum number for a
city population  $x_M$ can be established via consideration of
geographical peculiarities
as mountains \cite{aostamap} or oceans\cite{marmap}
 (See Fig.~\ref{fig8} for an example).
In such cases it is convenient to employ the transform $u =
\log_q(x/x_0)$. An associated, valuable macroscopic parameter is
$u_M = \log_q(x_M/x_0)$. We will be dealing then with a
``volume" $0<u<u_M$.\\

$\bullet$  {\it Total population $N_T$.}
We have $N_T=\sum_{i=1}^n\,x_i$ that gets transformed into
$N_T=x_0\sum_{i=1}^n\,e_q^{u_i}$. A useful quantity becomes
then $N=N_T/x_0$.\\

$\bullet$  {\it Total variance of $\dot{u}.$}
With reference to the dynamics, a useful observable is the total variance
for relative growth
$\sigma^2=\sum_{i=1}^n\langle(\dot{u}_i-\langle\dot{u}_i\rangle_t)^2\rangle/n$.
For a Gaussian form (see Fig.~\ref{fig1}) this quantity measures fluctuation-intensities.
Generalizing, this quantity can be defined by the covariance matrix with elements
$Q_{ij}=\langle(\dot{u}_i-\langle\dot{u}_i\rangle)(\dot{u}_j-\langle\dot{u}_j\rangle)\rangle$,
using its trace as a thermodynamical variable
$\mathrm{Tr}(Q)=\sum_{i=1}^nQ_{ii}$:
\ben
U &=& \frac{\tau}{2}\mathrm{Tr}(Q)\nonumber\\
  &=& \frac{\tau}{2}\sum_{i=1}^n\langle(\dot{u}_i-\langle\dot{u}_i\rangle)^2\rangle\nonumber\\
  &=& \frac{\tau}{2}n\sigma^2,
\een
where we add for dimensional convenience a factor $\tau/2$.\\

\subsection{Fundamental hypothesis for urban-thermodynamics}

\nd  Let us discuss the pertinent three hypothesis that we need
in our Scheme:\\

$\bullet$ \emph{H-I. Microscopic hypothesis.}\\

\nd We adopt as fundamental dynamical equation Eq. (\ref{dyn}) [$\dot{x}=kx^q$] for the population
of a center, linearized via the variable $u=\log_q(x/x_0)$. We will
think of the pair $(u,\dot{u})$ as constituting our social phase space
coordinates. We can speak of an:\\

$\bullet$ \emph{H-II. A priori phase space equiprobability in $(u,\dot{u})$.\cite{katz}}\\

\nd The probability density distribution for the $i-$th phase space cell centered
at $(u_i,\dot{u}_i)$ of size $dud\dot{u}$ is defined as
$\rho[\{(u_i,\dot{u}_i)\}_{i=1}^n]d^nud^n\dot{u}$. Accordingly to H-II, the system's
entropy is written as
\be
S[\rho] = -\int d^nud^n\dot{u}~\rho[\{(u_i,\dot{u}_i)\}_{i=1}^n]\log\left[\rho[\{(u_i,\dot{u}_i)\}_{i=1}^n]\right].
\ee
Since none of our macroscopic observables is able to distinguish amongst population
nuclei, towns are thus indistinguishable. In this case, the useful distribution is
the one-body density $\rho(u,\dot{u})$ defined as
\be
\rho(u,\dot{u}) =\int d^{n-1}ud^{n-1}\dot{u}~\rho[\{(u_i,\dot{u}_i)\}_{i=1}^n],
\ee
and thus,
\be
S[\rho] = -\int dud\dot{u}~\rho(u,\dot{u})\log\left[\rho(u,\dot{u})\right].
\ee
Macroscopic observables are written in terms of the one-body density as
\ben
n &=& \int dud\dot{u}~\rho(u,\dot{u}),\\
N &=& \int dud\dot{u}~\rho(u,\dot{u})e_q(u),\\
U &=& \frac{\tau}{2}\int dud\dot{u}~\rho(u,\dot{u})\dot{u}^2.
\een

$\bullet$ \emph{H-III. Maximum entropy principle (MaxEnt).\cite{katz}}\\

\nd Equilibrium is determined via constrained entropic maximization
using $n$, $u_M$, $N$ y $U$. This determines the equilibrium
density $\rho(u,\dot{u})$ that is a solution of the entropic
variational problem
\be\label{en}
\delta\left\{S[\rho]-\beta A[\rho]\right\} = 0,
\ee
with
\be
A = U - \mu n + p u_M + \Lambda N,
\ee
where $\beta$, $\mu$, $p$ and $\Lambda$ stand for the pertinent Lagrange multipliers,
that will be seen below to acquire the character of intensive thermal-quantities.

\subsection{Thermodynamical relations}

\nd We enter step 6 by considering the \emph{Lagrangian} $A[\rho]$ [and Lagrangian
density $a(u,\dot{u})$]. It reads
\ben
A[\rho] &=& \int dud\dot{u}~\rho(u,\dot{u}) a(u,\dot{u})\\
        &=& \int dud\dot{u}~\rho(u,\dot{u})\left\{
\frac{\tau}{2}\dot{u}^2 - \mu+ p v(u) + \Lambda e_q(u)\nonumber
\right\},
\een
where the volume condition is enforced by an infinite-well potential
\be
v(u)=\left\{
\begin{array}{ll}
u_M/n  &for\,\,\,0<u<u_M;\\
\infty &otherwise
\end{array}
\right.
\ee
The well-known general solution to the entropic problem Eq.~(\ref{en}) is\cite{katz}
$\rho(u,\dot{u}) = \exp[-\beta a(u,\dot{u})]$, so that
\be
\rho(u,\dot{u}) = \frac{n}{Z}e^{-\frac{\beta\tau}{2}\dot{u}^2-\beta\Lambda e_q(u)}~~(0<u<u_M),
\ee
where the normalization factor $Z$ (partition function) becomes
\ben
 Z &=& \int_{-\infty}^{\infty} d\dot{u}\int_0^{u_M} du~e^{-\frac{\beta\tau}{2}\dot{u}^2-\beta\Lambda e_q(u)}\nonumber\\
   &=& \sqrt{\frac{2\pi}{\beta\tau}}E_q(\beta\Lambda,u_M),
\een
with $E_q(l,m)$ the generalized exponential function of order $q$
\be
E_q(l,m) = E_q(l) - e^{(1-q)m}E_q(le^m).
\ee
Our constraints in $U$ and $N$ determine the multipliers $\beta$ and $\Lambda$-values.
On the one hand, we have the $\dot{u}-$variance
\ben\label{sigma}
 U &=& \frac{\tau}{2}n\int_{-\infty}^{\infty} d\dot{u}~\sqrt{\frac{\beta\tau}{2\pi}}e^{-\frac{\beta\tau}{2}\dot{u}^2}\dot{u}^2\nonumber\\
   &=& \frac{n}{2\beta},
\een
and on the other one, we have to deal with the total population
\ben\label{N}
 N &=& n \int_0^{u_M} du~\frac{e_q(u)e^{-\beta\Lambda e_q(u)}}{E_q(\beta\Lambda,u_M)}\nonumber\\
   &=& n \frac{E_{q-1}(\beta\Lambda,u_M)}{E_q(\beta\Lambda,u_M)},\nonumber\\
   &=& n F_q(\beta\Lambda,u_M),
\een
where we use the function $F_q(l,m)=\partial_l\log[E_q(l,m)]$. We obtain from the former the
direct result
\fbox{\parbox{\linewidth}{
\be\label{sigma2}
\beta =  \frac{n}{2U},
\ee
}}
and, via inversion of the latter equation
[defining first $L_q(f,m)=F_q^{-1}(f,m)$ and thus $F_q[L_q(f,m),m]=f$], we finally obtain the relation
between the system variables \emph{(equation of state)}
\fbox{\parbox{\linewidth}{
\be\label{ee}
\beta\Lambda =  L_q(N/n,u_M).
\ee
}}

\nd Note that we have intensive quantities on the left hand
side, while extensive ones appear in the r.h.s..  The
entropy becomes
\ben
 S &=& n\log\left[\frac{1}{n}\sqrt{\frac{2\pi}{\beta\tau}}E_q(\beta\Lambda,u_M)\right]\nonumber\\
    && + n\left[\frac{1}{2}+\beta\Lambda F_q(\beta\Lambda,u_M)\right].
\een
Using now Eq. (\ref{ee}) we can recast things in term of
the natural variables as
\ben
 S(U,n,u_M,N) &=& n\log\left[\frac{2}{n}\sqrt{\frac{\pi U}{n\tau}}E_q[L_q(N/n,u_M),u_M]\right]\nonumber\\
               && + \frac{n}{2} + L_q(N/n,u_M) N.
\een
It is easy to verify, but crucial to our present goals,  that
macroscopic observables and Lagrange multipliers become linked
entropic-wise via
\ben
\beta   &=&                \left.\frac{\partial S}{\partial U}\right|_{n,u_M,N},\\
\mu     &=& \frac{1}{\beta}\left.\frac{\partial S}{\partial n}\right|_{U,u_M,N},\\
p       &=& \frac{1}{\beta}\left.\frac{\partial S}{\partial u_M}\right|_{U,n,N},\\
\Lambda &=& \frac{1}{\beta}\left.\frac{\partial S}{\partial N}\right|_{U,u_M,n}.
\een
The first relation leads to Eq. (\ref{sigma2}), also showing that the $\beta-$multiplier
is the inverse of the variance $\beta=1/\tau\sigma^2$. The last relation takes us to
Eq. (\ref{ee}) and is indeed one of our equations of state. The other two are
\nd\fbox{\parbox{\linewidth}{
\ben
\beta\mu &=& \log\left[\frac{2}{n}\sqrt{\frac{\pi U}{n\tau}}E_q[L_q(N/n,u_M),u_M]\right] - 1\nonumber\\
          && - 2L_q^{(1,0)}(N/n,u_M) \left(\frac{N}{n}\right)^2
\een
}}
and

\nd\fbox{\parbox{\linewidth}{
\ben\label{p}
\beta p &=& n\frac{\exp[-L_q(N/n,u_M)e_q(u_M)]}{E_q[L_q(N/n,u_M),u_M]}\nonumber\\
         && + 2L_q^{(0,1)}(N/n,u_M) N.
\een
}}

\nd 
At this stage the reader will agree that it is fair to assert that
our goal has been successfully reached. \emph{We have indeed constructed a social
thermodynamics for urban population flows}. The following equivalence may be established
vis-a vis  the thermodynamics of chemical species:

\nd\fbox{\parbox{0.97\linewidth}{
\begin{itemize}
  \item Temperature     $\leftrightarrow$ Variance of relative growth.
  \item Number of inhabitants   $\leftrightarrow$ Number of particles.
  \item Number of towns  $\leftrightarrow$ Number of chemical species.
  \item Volume   $\leftrightarrow$ Maximum possible town's population.
\end{itemize}
}}

\section{Application: the scale-free ideal gas (SFIG)}

\nd This is our final step 7. We envision two main regimes,
according to the $\Lambda-$value: $\Lambda=0$ and $\Lambda>0$.

\subsection{The SFIG-in-a-box}

\nd We will consider in some detail the first case here. Different scenarios can be
associated to $\Lambda\rightarrow0$: i) the system is not isolated and exchanges
population with its surroundings, with a maximum-size constraint, ii) the triplet
$n$, $N$, $u_M$ is such that the equation of state yields $\Lambda=0$, i.e.,
$N/n = \log_{q-1}[e_q(u_M)]/u_M$, or iii) no size-limitation exists
($u_M\rightarrow\infty$) but $N/n$ is large enough to consider $\Lambda\sim0$.
In the latter case one can obtain an effective $u_M-$value from normalization
such that $u_M=E_q(\beta\Lambda)$. When $\Lambda=0$ the Lagrangian $A$ is written as
\be
A = U - \mu n + p u_M,
\ee
so that we do not need knowledge of $N$. The equilibrium density is
\be
\rho(u,\dot{u})dud\dot{u} = \frac{n}{u_M}\sqrt{\frac{\beta\tau}{2\pi}}~e^{-\frac{\beta\tau}{2}\dot{u}^2}dud\dot{u}~~(0<u<u_M).
\ee
The partial density $\rho(u) = \int d\dot{u}\rho(u,\dot{u})=n/u_M$
is constant in $u$ so that $x$ is given by a power-law
\be
 \rho_X(x)dx = \rho[u(x)]\frac{du}{dx}dx = n\frac{x_0^{q-1}}{u_M}\frac{dx}{x^q},
\ee
with an associated rank-plot given by
\be\label{xr}
x(r) = x_0e_q[u_M(1-r/n)],
\ee
where $r$ is the rank from 1 to $n$.
Comparing with the equilibrium densities found above in our numerical
experiments with random walkers, a very nice fit ensues as seen in Fig.~5, which \emph{validates our
methodology}. The entropy becomes
\be
S(U,n,u_M) = n\log\left[\frac{2}{n}\sqrt{\frac{\pi U}{n\tau}}u_M\right] + \frac{n}{2},
\ee
resembling that of the one-dimensional ideal gas. The state-equations are
\be
\beta\mu = \log\left[\frac{2}{n}\sqrt{\frac{\pi U}{n\tau}}u_M\right] - 1,
\ee
and
\be
\beta p = \frac{n}{u_M},
\ee
in exact agreement with the ideal gas scenario.

As empirical $q=1$-examples we show the cases of i) Marshall Islands,\cite{mar} ii) d'Agosta Valley
(Italy) \cite{ita} and iii) Huelva-province (spain)\cite{ine}.
In all instances the relative growth is nearly independent of the
population (with some finite-size noise) as in the case of Huelva,
or with a secondary constant trend for low-populated cities, as in
d'Agosta Valley's intance. Thus we  consider that the microscopic
dynamics fits the proportional growth hypothesis, with  densities
$\rho(u,\dot{u})$ nicely adapted to the ensuing thermodynamic
predictions.
In all cases,  geographical conditions set strong limits to the
city-sizes. The values for the macroscopic parameters are shown in Table~\ref{table1}.
Remarkably enough, the pressure $p$ due to the limited space is  highest
for the Marshall Islands. Indeed, this system exhibits the lowest volume $u_M$ and the
lowest $\beta$ (highest ``temperature'') for a large number of units $n$.

\begin{table}[ht]
\begin{tabular}{c|c c c}
   & Marshall Islands & Agosta Valley & Huelva \\
\hline\hline
$n$     & $160$ & $74$    & $79$ \\
$x_0$   & $2.14$ & $126$   & $206$ \\
$u_M$   & $0.038$ & $0.048$ & $0.059$ \\
$\beta$ & $0.504$ & $2.05$  & $11.7$ \\
$p$     & $0.829$ & $0.0133$  & $0.0115$ \\
\end{tabular}
\caption{\label{table1} Macroscopic parameters for the four SFIG-in-a-box examples ($\tau=100$ years$^2$).}
\end{table}

\subsection{The SFIG under total population-constraint}

\nd We now consider $\Lambda>0$, with $u_M\rightarrow\infty$ for simplicity.
This case describes regions where internal migration dominates the microscopic dynamics,
and no upper limit is found for the city-size. According to the equation of state
Eq.~(\ref{p}), $p=0$ in this limit, so that we deal with the lagrangian
\be
A = U - \mu n + \Lambda N.
\ee
The equilibrium density is
\be
\rho(u,\dot{u})dud\dot{u} = \frac{n}{E_q(\beta\Lambda)}\sqrt{\frac{\beta\tau}{2\pi}}e^{-\frac{\beta\tau}{2}\dot{u}^2-\beta\Lambda e_q(u)}dud\dot{u}~~(0<u),
\ee
and the ecuation of state can be written in the form
\be
 N = n \frac{E_{q-1}(\beta\Lambda)}{E_q(\beta\Lambda)}.
\ee
The partial density for $x$ is given by a power-law with exponential cut-off
\be
 \rho_X[x]dx = n\frac{x_0^{q-1}}{E_q(\beta\Lambda)}\frac{e^{-\beta\Lambda x}}{x^q}dx,
\ee
with an associated rank-plot
\be\label{rpq}
x(r) = \frac{x_0}{\Lambda}E^{-1}_q\left[E_q(\Lambda)r/n\right].
\ee
Again, this result fits the numerical equilibrium densities found above in
our numerical simulations (Fig.~6), \emph{validating again our methodology}.

\nd This is the most common situation in the Spanish provinces
(for more details see Ref.~\cite{emp}). We have found  nice agreement between the $q$
value obtained from a fit to the microscopic dynamics and the $q-$value obtained
from the fit of the rank-plot to Eq.~(\ref{rpq}). We show some examples in
Fig. \ref{fig10}, and the associated macroscopic  numerical results  in Table~\ref{table2}.
In the examples presented below, using the parameter $\Lambda$ as a measure of the pressure generated by the total
population constraint, it turns out that Alicante is the province with the highest pressure and Girona
that with the lowest one,
correlated with a highest and a lowest `temperature',
respectively.

\begin{table}[ht]
\begin{tabular}{c|c c c c}
   & Alicante & Almer\'ia & Girona & Lleida \\
\hline\hline
$n$                  & $140$          & $101$          & $220$          & $230$\\
$x_0$                & $83.9$         & $147$          & $141$          & $105$\\
$N$                  & $18355.7$      & $3245.53$      & $4597.37$      & $2818.12$\\
$\log(\beta\Lambda)$ & $-6.26$        & $-5.35$        & $-5.18$        & $-4.02$\\
$\Lambda$            & $4.03~10^{-3}$ & $1.83~10^{-4}$ & $7.89~10^{-5}$ & $4.1~10^{-4}$\\
$\beta$              & $0.47$         & $25.9$         & $71.1$         & $43.8$\\
$q$                  & $0.862$        & $1.135$        & $1.27$         & $1.16$\\
\hline
   & Navarra & Vizcaya & Zaragoza & Granada \\
\hline\hline
$n$                  & $271$          & $111$          & $292$          & $167$\\
$x_0$                & $47.6$         & $219$          & $40.7$         & $263$\\
$N$                  & $8976.55$      & $3571.01$      & $6906.03$      & $2539.5$\\
$\log(\beta\Lambda)$ & $-5.16$        & $-5.18$        & $-5.06$        & $-4.01$\\
$\Lambda$            & $6.69~10^{-4}$ & $1.18~10^{-4}$ & $6.22~10^{-4}$ & $2.03~10^{-3}$\\
$\beta$              & $8.57$         & $47.4$         & $10.2$         & $8.91$\\
$q$                  & $1.06$         & $1.08$         & $1.18$         & $1.02$\\
\end{tabular}
\caption{\label{table2} Macroscopic parameters for the SFIG-under-pop-constraint examples.}
\end{table}

\section{Conclusions}

\nd  After initially introducing some useful social-macroscopic and
social-stochastic quantities we have
\begin{enumerate}
  \item  Postulated social, dynamic microscopic equations.
  \item  Validated them  using urban population data.
  \item  Performed  numerical simulations with random walkers that conclusively
         demonstrated  that a description using many microscopic variables
         has as a counterpart a macroscopic one with few parameters.
  \item  Showed  that such macroscopic description can be given an appropriate
         MaxEnt form after constructing a ``social" phase space, that allows
         one to derive thermodynamic-like relations amongst our macro-parameters.
  \item  Finally, as an application, we successfully analyzed  urban flows as
         modelled by a scale invariant ideal gas.
\end{enumerate}

\begin{figure}
\includegraphics[width=0.45\textwidth]{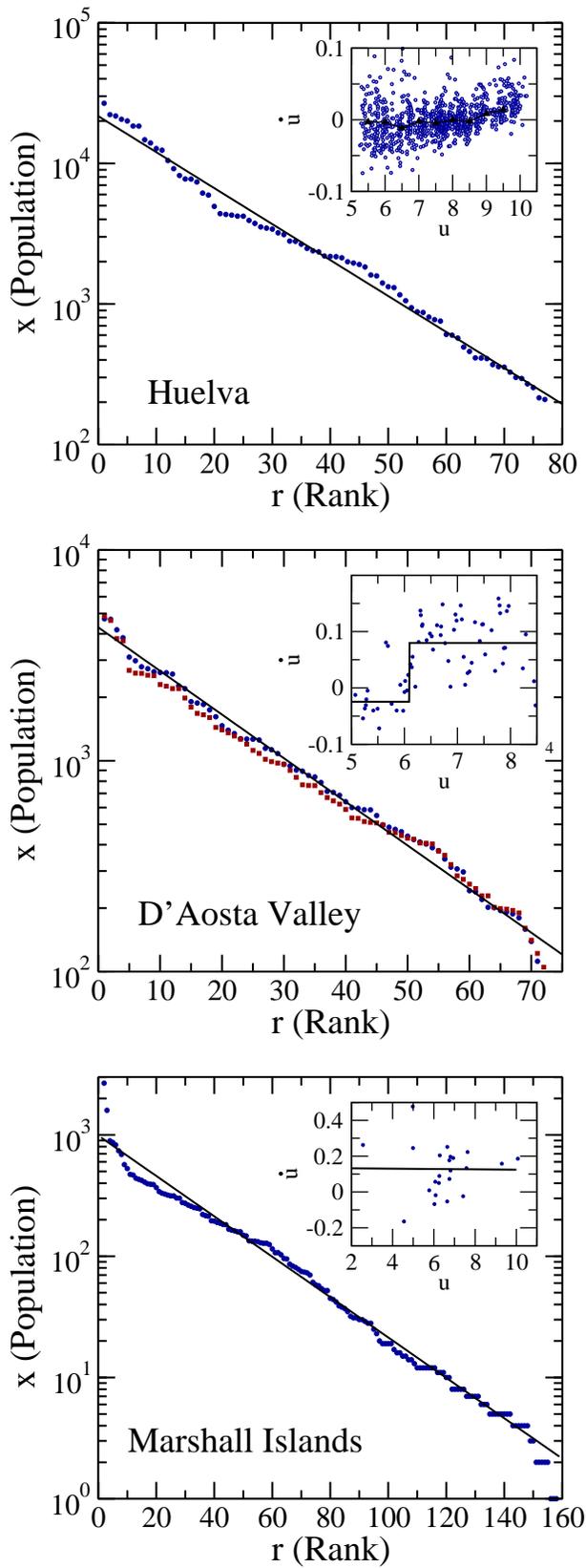}
\caption[]{SFIG-in-a-box examples, from top to bottom: RD of Huelva-province (Spain),
D'Aosta Valley (Italy), and Marshall Islands (dots), compared with distribution Eq. (\ref{xr})
(lines). In the insets, the relative growth $\dot{u}$ vs. the logarithmic population $u$.}\label{fig9}
\end{figure}
\begin{figure*}
\includegraphics[width=0.95\textwidth]{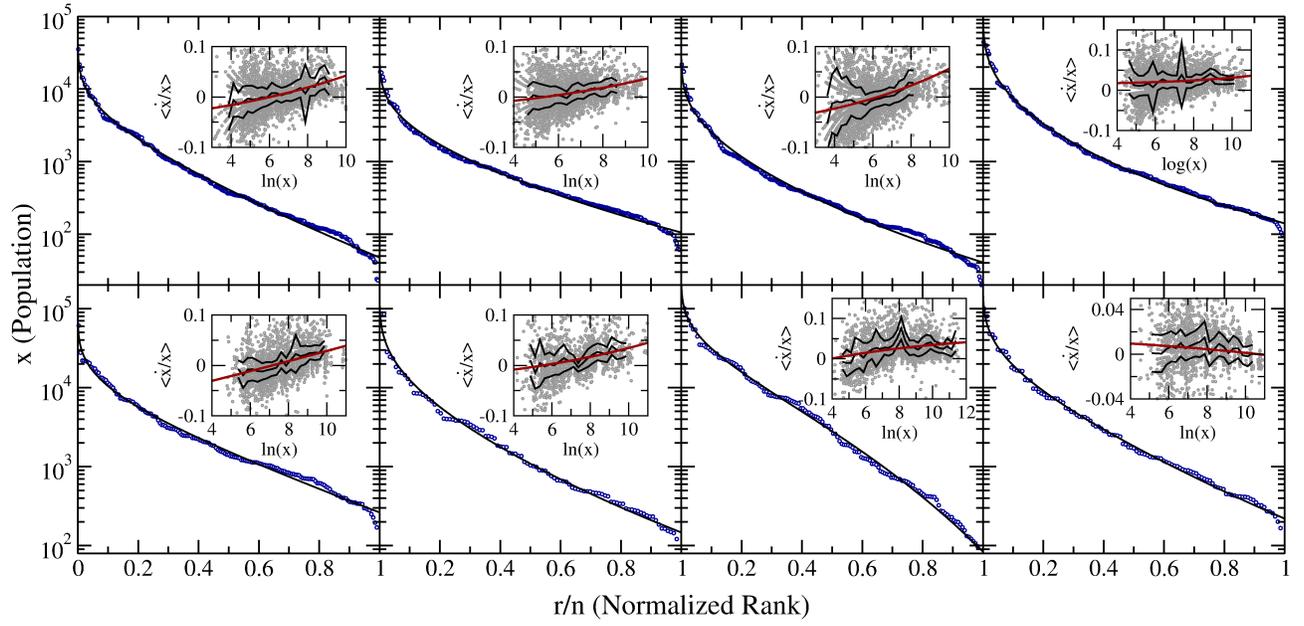}
\caption[]{SFIG under total-population constraint examples, in the reading order: RD of
Navarra, Lleida, Zaragoza, Girona,
Granada, Almer\'ia, Alicante, and Vizcaya
(dots) and RD of Eq. (\ref{rpq}) (lines). Insets: relative growth $\dot{x}/x$
vs. log-population $\log x$ fitted to Eq. (16) with the same value of $q$ as in the
RD.
}\label{fig10}
\end{figure*}

\end{document}